\documentclass{elsart}
\usepackage{natbib,psfig}
\begin{document}
\runauthor{K. Mannheim}
\begin{frontmatter}
\title{Neutrino Oscillations and Blazars}
\author{Karl Mannheim}{}
\address{
Universit{\"a}ts-Sternwarte G{\"o}ttingen, 
Geismarlandstra{\ss}e 11, D-37083 G{\"o}ttingen}
\begin{abstract}

Three independent predictions follow from postulating the existence of protons
co-accelerated with electrons in extragalactic jets (i) multi-TeV gamma ray
emission from nearby blazars, (ii) extragalactic cosmic ray protons up to $\sim
10^{20}$~eV, and (iii) extragalactic neutrinos up to $\sim 5\times 10^{18}$~eV.
Recent gamma ray observations of Mrk~421 and Mrk~501 employing the air-Cerenkov
technique are consistent with the predicted gamma ray spectrum, if one corrects
for pair attenuation on the infrared background.  Prediction (ii) is consistent
with cosmic ray data, if one requires that jets are responsible for a
at least a sizable fraction of the
extragalactic gamma ray background.  With kubic kilometer neutrino
telescopes, it will be possible to test (iii), although the muon event rates
are rather low. Neutrino oscillations can increase the event rate by inducing
tau-cascades removing the so-called Earth shadowing effect.

\end{abstract}
\end{frontmatter}

\section{Introduction}

In the early days after the discovery of extragalactic radio sources, it was a
widely held belief that the relativistic electrons responsible for the observed
synchrotron emission are secondary electrons from pp-interactions of
accelerated protons on ambient gas \cite{B56,P69}.  Protons and ions were known to be the by
far dominant species in the observed cosmic rays, and this was expected to be
mirrored at their acceleration sites.  Nevertheless, the picture was soon given
up, since it would require enormous amounts of target matter in the jets which
is inconsistent with plausible
energetics and the non-observation of emission lines
and bremsstrahlung.  An electron-positron composition of the jets was also
suggested from theory claiming that the plasma feeding radio jets should be due
to pair production in the ergosphere of a maximally rotating black hole 
\cite{BZ77}.  The
dielectric properties of radio jets inferred from observations of
weak Faraday rotation and circular
polarization in the jet of 3C279 seem to lend further support to a light
composition \cite{W98}.  However, this picture also bears several fallacies.

Target matter does not have to be present to obtain efficient cooling
of accelerated protons.  Photo-production of secondary
particles and synchrotron emission can become important, if the proton
energy is high enough, since
the cooling rate
for these processes decreases with energy.
In fact, in a
statistical acceleration process in which there is enough time and space
to balance
acceleration energy gains against energy losses, it is an inevitable
consequence that the protons reach ultrahigh energies
\cite{BS87,S87}. 
Even if the original composition of the jet
plasma were light, polluting baryons from the ambient medium 
would quickly take over most
of the jet's momentum, so that the acceleration mechanism must eventually tap
baryonic kinetic energy.  In fact, for gamma ray bursts, this is believed to be
the crucial explanation for the fact that the burst energy is tapped only by
shock fronts far away from the site of the original pair fireball \cite{MR93}. 
Therefore, it is the most natural assumption that the observed
nonthermal emission is partly due to the accelerated electrons, and partly
due to the accelerated protons \cite{MKB91}.
Inferences of the plasma composition based on 
measurements of polarization and Faraday rotation
are based on the assumption that
the electron distribution observed in the optically thin synchrotron regime
traces down to lower energies which is known to lead to theoretical
inconsistencies \cite{MC69}.  Since the radiative properties are determined by the
particles with the highest energies but the dielectric properties by those
with the lowest energies, there is a general mismatch in the conclusions drawn
from observations sensitive to either regime and one must be careful with
claims about jet composition.

It may
be a good advice not be too narrow-minded when confronted with
a new observational result such as
multi-TeV emission from blazars
and to include as many independent
facts as possible in its interpretation.
If the gamma rays were indeed primarily
of a hadronic origin, there are a number
of corollaries which allow to falsify the claim, whereas
the arguments for a purely leptonic origin of the gamma rays boil down
to some version of Ockham's razor (``it is more economic to use the
observed electrons to model the gamma ray emission'').
I wonder whether this is enough to
get around the symptomatic fact that there was no prediction of multi-TeV emission from
leptonic models prior to the observations.  Moreover, there are
problems with leptonic models, 
such as the missing intrinsic curvature in the multi-TeV
spectrum of Mrk~501 and the surprisingly low magnetic field
values as was pointed out in ref. \cite{M98}.  
Although the most recent HEGRA spectra of Mrk~501
\cite{Konol99} do show some curvature,
the lower limits
on the infrared background imply that the intrinsic spectrum
must be rather flat or even up-turning  \cite{DJD99}.  
 
One might as well argue that it is much
more in the sense of Ockham's razor (i.e. more
economic),
if one finds that the same sources which very
likely produce most of the extragalactic
gamma rays would at the same time produce 
cosmic rays at ultrahigh energies 
where it is difficult to find any other astrophysical source
supplying enough power.
There are two independent observations and only one model.
According to Landau, a third independent fact is needed to consider
a theoretical claim seriously.  Here I consider
the high-energy neutrino emission 
intimately connected
with multi-TeV gamma rays from hadronic accelerators as the
missing piece of information.
To first elucidate the connection between extragalactic high-energy emissions
in the framework of global energetics,
Sect.2 quantifies the non-thermal energy that may be released by active
galactic nuclei and their jets integrated over their cosmic history.
This qualifies extragalactic jets as possible sources of the
ultrahigh energy cosmic rays and implies multi-TeV emission
from the jets due to proton energy losses at their acceleration site.
Section 3 discusses the predicted
spectra from a simple quasi-stationary unsaturated synchrotron
cascade emission model.  Bearing on the assumption of a strong
evolution of their luminosity density, the neutrino and cosmic ray
spectra from extragalactic jets using the assumptions of
the hadronic gamma ray emission model are computed in Sect.4.  
Since the expected peak of the
neutrino spectrum lies at energies in excess of 100~TeV at which
the Earth becomes optically thick with respect to neutrino absorption,
Sect.5 extends the discussion by including the profound effects of neutrino
oscillations.

\section{Origins of extragalactic background radiation}

\begin{figure*}
\centerline{\psfig{figure=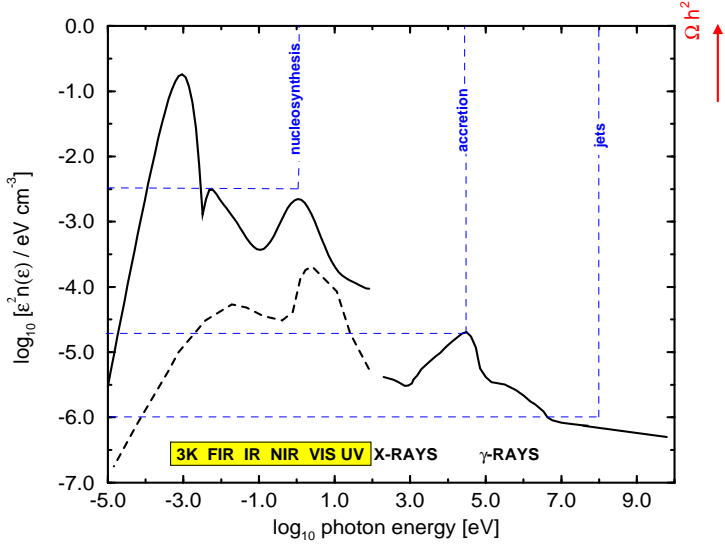,width=10cm,height=8cm}}
\caption{Sketch of the present-day energy density
of the extragalactic radiation background from radio waves to
gamma rays. }
\label{fig1}
\end{figure*}

Inspection of Fig.~\ref{fig1} shows an interesting pattern
in the present-day energy
density of the diffuse isotropic background radiation 
consisting of a sequence of bumps each with a strength that is
decreasing with photon energy.
The microwave bump is recognized as the signature of the big bang
at the time of decoupling with its energy density given by
the Stefan-Boltzmann law $u_{\rm 3K}=\sigma T^4$.
The bump in the far-infrared is due to star formation in early galaxies, 
since part of the stellar light, which appears
as the bump at visible wavelengths, is reprocessed by
dust obscuring the star-forming regions.  The energy density of the two bumps
can be related to the present-day heavy element abundances.
Heavy elements with present-day mass fraction 
$Z=0.03$ were produced in early bursts of star formation
by nucleosynthesis with radiative efficiency 
$\epsilon=0.007$ yielding the present-day energy density
\begin{equation}
u_{\rm ns}\sim {\rho_* Z \epsilon c^2\over 1+z_{\rm f}}
\end{equation}
where $\rho_*$ denotes the mass density of baryonic matter
and $z_{\rm f}$ the formation redshift corresponding to
the era of maximum star formation.  This is, of course, only
a very rough approximation of the true star formation history,
but good enough to set the scale for an argument pertaining
to the global energetics.  In particular, the ratio between
the energy released by stars and by other sources with the
same formation history is independent
of its details.
Let $\Omega_*$ denote
the baryon density in terms of the critical density
of the Universe and
$h=H_\circ/100~$km~s$^{-1}$~Mpc$^{-1}$ the
dimensionless Hubble constant, then
the energy density takes the value
\begin{equation}
u_{\rm ns}\sim
6\times 10^{-3}\left(\Omega_*h^2\over 0.01\right)\left(1+z_{\rm f}\over
4\right)^{-1}\ {\rm eV~cm^{-3}}
\end{equation}
and should represent the energy density of
the sum of the far-infrared and optical bumps.
Probably all galaxies (except dwarfs) contain
supermassive black holes in their centers which are actively
accreting over a fraction of $t_{\rm agn}/ t_*\sim 10^{-2}$
of their lifetime implying that the electromagnetic radiation
released by the accreting black holes amounts to
\begin{equation}
u_{\rm accr}\sim {\epsilon_{\rm accr}M_{\rm bh}
\over  Z\epsilon M_*}
{t_{\rm agn}\over  t_*}u_{\rm ns}\sim 1.4\times 10^{-4}\ \rm eV~cm^{-3}
\end{equation}
adopting the accretion efficiency $\epsilon_{\rm accr}=0.1$
and the black hole mass fraction $M_{\rm bh}/M_*=0.005$
\cite{R98}.  Most of the accretion power emerges in
the ultraviolet where the diffuse background is unobservable
owing to photoelectric absorption by the neutral component of the
interstellar medium.  However, a fraction of $u_{\rm x}/u_{\rm bh}\sim
20\%$ taken from the average quasar spectral energy distribution
\cite{S89} shows up in hard X-rays 
producing the diffuse isotropic X-ray background
bump with 
$u_{\rm x}\sim 2.8\times 10^{-5}~\rm eV~cm^{-3}$
\cite{G92}.
Jets with non-thermal $\gamma$-ray emission show up 
only in the radio-loud fraction $\xi_{\rm rl}\sim 20\%$ of
all AGN and their kinetic power roughly equals
the accretion power \cite{RS91}.  Hence
one obtains for the energy density due to extragalactic jets
\begin{equation}
u_{\rm j}=\left(\xi_{\rm rl}\over 0.2\right)u_{\rm accr}\sim 
\left(\xi_{\rm rl}\over 0.2\right)2.8
\times 10^{-5}~\rm eV~cm^{-3}.
\end{equation}
This energy is released in relativistic particles, magnetic fields,
and pdV thermodynamic work against the ambient medium into which
the jets propagate.  Adopting a radiative efficiency of $\xi_{\rm rad}=10\%$
for the jets, the gamma ray (cosmic ray, neutrino)
energy released by the jets can amount to
a maximum present-day energy density of 
\begin{equation}
u_\gamma\sim \xi_{\rm rad}u_{\rm j}\sim 2.8\times 10^{-6}
\left(\xi_{\rm rad}\over 0.1\right)\left(\xi_{\rm rl}\over 0.2\right)
\ \rm eV~cm^{-3}
\end{equation}
which comes remarkably close to 
the energy density $3.2\times 10^{-6}$~eV~cm$^{-3}$
of the extragalactic gamma ray background observed
between 100~MeV and 30~GeV\footnote{
Note that the flux in the gamma ray background observed by CGRO is close
to the bolometric gamma ray flux,  since pair attenuation and cascading
must lead to a turnover of the background spectrum above $20-50$~GeV
for extragalactic source populations
\cite{MP96,Salamon98}}
using the spectrum given in ref.~\cite{Sreekumar}.
Note that this is consistent with having a particle acceleration
efficiency in radio jets
which is of the same order of magnitude as the 13\% efficiency required for
supernova remnants to produce the Galactic cosmic
rays.

Protons can achieve a high radiative efficiency due to photo-production
of secondary particles or synchrotron emission only if they reach
ultrahigh energies.  This can be seen most easily for synchrotron radiation
in which case the proton Lorentz factor must be larger 
than the electron Lorentz factor by $(m_p/m_e)^3$
to produce the same cooling rate. Particles with energies exceeding
$6\times 10^{18}$~eV may seem outrageously exotic to many astrophysicists, but
they are actually observed in the local cosmic ray spectrum.
These particles cannot be confined by the Milky Way, but nevertheless
show a near-isotropic sky distribution.
Furthermore, the local cosmic ray spectrum flattens above 
$6\times 10^{18}$~eV indicating a separate extragalactic source population.
It will be argued in Sect.4 that only an
extragalactic source population as strong as the one supplying
the diffuse isotropic gamma rays can be considered for their origin.
Before highlighting these arguments,
it is shown in the next section that protons at such high energies
produce interesting gamma ray spectra
owing to the photo-production of secondaries.  

\section{Proton blazar predictions and observed multi-TeV
spectra}

A large number of emission models for radio jets based on a shock-in-jet
scenario exist and have been shown to explain most of the low-frequency jet
phenomenology using a number of parameters such as jet kinetic energy,
magnetic field gradient, opening angle of the flow channel, etc. within
plausible ranges.  It is straightforward to include relativistic protons
in such models assuming they have a spectrum with the same slope
but different maximum and minimum
energies (due to different energy losses, gyro-resonant thresholds, and
Larmor radii).  Among the interesting consequences of the protons
are (i) a higher nonthermal pressure and (ii) more gamma ray flux
(adding to the Compton flux from the accelerated electrons).
The equipartition magnetic field
strength increases according to
$B\propto \kappa^{2/7}$ where $\kappa=u_p/u_e$ denotes the ratio
between the energy density in relativistic protons and electrons,
respectively, and this may help to alleviate some problems
for pair jets such as 
the observed pressure support of the radio lobes 
in NGC 1275 against the surrounding hot intracluster medium
of the Perseus cluster
\cite{Boehringer}.  

\begin{figure}
\centerline{\psfig{figure=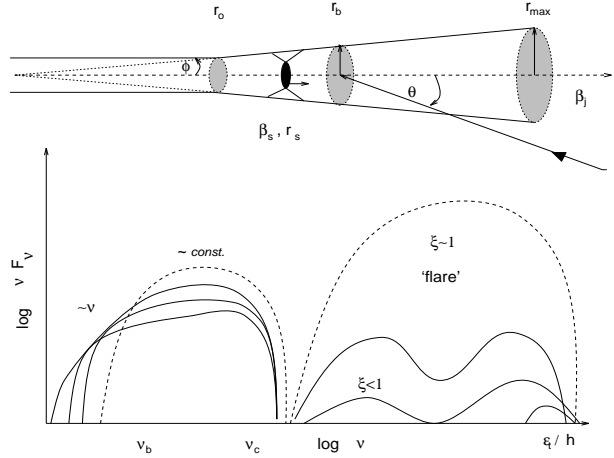,width=8cm,height=6cm}}
\caption{Conical geometry 
for the relativistic
jet with Lorentz factor $\gamma_{\rm j}$ and
opening angle $\Phi\sim 1/2\gamma_{\rm j}$ 
assumed in the proton blazar model.
Note that the total proton-initiated cascade emission from the cone
is dominated by
the emission from the region of highest surface brightness temperature
in the submm-infrared regime at $r=r_{\rm b}$
and is therefore more susceptible
to variations due to traveling shocks (at position $r_{\rm s}$
with velocity $\beta_{\rm s}$). The sketch of the spectral energy
distribution
indicates the flux contributions from various scales $r$ in the jet.
An additional effect increasing the gamma ray flux from the
region of highest surface brightness temperature is indicated
by the parameter $\xi$ which is proportional to the proton maximum
energy and which is expected to decrease with $r$ due to the
nonlinear development of the shock structure.}
\label{fig2}
\end{figure}

In terms of simplicity and predictive power, the original
Blandford \& K\"onigl conical jet model \cite{BK79}
is useful and has been investigated for
the radiative signatures of the protons \cite{MKB91}.
The model describes the stationary emission from a conical section of
a free relativistic jet 
and is therefore of limited applicability to
non-stationary features in the observed spectra.  
The relativistic particle and magnetic energy
density decreases as $\propto r^{-2}$ along the jet, i.e. the jet
is isothermal.
The emitting conical section of the jet may be 
thought of as the unresolved
superposition of
a number of shocks traveling down the jet, and shock acceleration
keeps the nonthermal energy constant. 
Such a jet emits a flat
radio spectrum up to the frequency
where the spectrum steepens by one power due to the energy losses, typically
in the
submm-infrared regime.  The flux at this break frequency
is dominated by the jet cone near its apex.
The threshold frequency $\nu_{\rm th}$
for the production of pions in head-on
collisions between photons and protons with Lorentz factor $\gamma_p$
is given by
\begin{equation}
\nu_{\rm th}\simeq 10^{12}
\left({\gamma_p\over 10^{11}}\right)^{-1}\ \rm Hz
\end{equation}
showing that the region of maximum surface brightness temperature
is most important for the cooling of protons.  Owing to electromagnetic
cascading the electromagnetic power injected into the jet plasma
by the cooling baryons is redistributed smoothly
over the X-ray and gamma ray bands.
Therefore, the emission component with
the largest energy flux dominates
the entire high-frequency spectrum and proton-initiated
cascade spectra 
have been computed only for this zone.
An integration along the jet axis
(which is necessary to obtain the
flat radio spectrum)
would only lead to marginal corrections in the gamma ray regime.
The geometry is depicted in Fig.~\ref{fig2} together with a sketch of 
the emission components from the various scales.
Due to the small size of the region dominating the gamma ray
energy flux compared with the total volume of the conical jet,
traveling shocks would make the gamma ray emission much more
susceptible to flux variability than the radio-infrared emission
in qualitative agreement with the observations.
As long as the perturbation time scale is larger than
the proton acceleration time scale, the emission spectra can still
be computed as quasi-stationary spectra, and it is an open theoretical
challenge to solve for the spectra in the general time-dependent case.
An interesting time-dependent solution exists for a simplified version
of Compton-scattering dominated proton-initiated cascades
\cite{KirkMast}.  Using the Doppler factor $\delta$ to convert
between comoving and observer's frame, the proton
acceleration time scale assuming Bohm diffusion
is given by
\begin{equation}
t_{\rm acc}\simeq {r_{\rm g}\over \delta c}={\gamma_p m_p c\over \delta
e B}\simeq
3.5\times 10^4 \left(\gamma_p\over 10^{11}\right)
\left(B\over 30~{\rm G}\right)^{-1}\left(\delta\over 10\right)^{-1}~ \rm s
\end{equation}
in comfortable agreement with some of
the observed TeV flux variation time scales.
The comoving frame gyro-radius of the protons 
\begin{equation}
r_{\rm g}\simeq
10^{16}\left(B\over 30~{\rm G}\right)^{-1}\left(\gamma_p\over
10^{11}\right)\ {\rm cm}
\end{equation}
satisfies the
constraint $r_{\rm g}\le r_{\rm j}$ if the jet radius is the
lightcylinder radius of an MHD jet from a black hole of mass $M$
and Schwarzschild radius $r_{\rm S}$ in which case
$r_{\rm j}\simeq 
100 r_{\rm S}\simeq 3\times
10^{16}(M/10^9M_\odot)$~cm.
For time scales longer than the
proton acceleration time scale,
correlated flux variations from hard X-rays to TeV gamma rays
are expected as a typical phenomenon (within the model assumptions).
If the hard X-ray emission or the gamma ray emission in the EGRET
band is dominated by 
emission from the accelerated electrons, 
the TeV variability can be different.  This
case would argue for a higher ratio of the photon to magnetic
energy density $u_\gamma/u_B\sim 1$ than in the generally
assumed case $u_\gamma/u_B<1$ in the proton blazar model.
Note also that variability
on shorter time scales must be expected under realistic
conditions, e.g. due to the presence of inhomogeneities along the jet. 
These variations can occur on time scales down
to the cooling time scale of the
cascade electrons, and require the solution of the time-dependent
cascade equations.  The short-term
variability behavior of the cascades is
very complicated owing to the pair production threshold 
varying with gamma ray energy.  

The TeV detections of nearby blazars confirm the prediction of
the proton blazar model that the relativistic jets should
be optically thin to gamma ray emission below $\sim 1$~TeV. 
This
means that the conversion of injected gamma rays into pairs
and vice versa becomes unimportant below TeV
(unsaturated cascades).  Emission above $\sim 1$~TeV is expected
to be optically thick, but nevertheless of approximate power law
shape.
Synchrotron cooling is assumed to be the dominant process replenishing
the gamma rays from the pairs.  For the entire parameter space,
synchrotron emission remains non-relativistic so that
the characteristic synchrotron photon energy is much less
than that of the radiating electrons assuring rapid convergence
of the cascade equations (which can be brought into the form
of a Volterra integral equation of the second kind)
by Banach's fixed point theorem.  
The shape of the multi-TeV spectrum follows from very
simple considerations.
It is assumed that the protons have a differential distribution
$dN/dE\propto E^{\rm -s_{\rm p}}$.  The electron distribution has the 
same slope in the optically thin range, but in the energy range
responsible for producing the target photons for the protons
their spectrum is steeper by one power $s_{\rm e}=s_{\rm p}+1$ owing
to energy losses.  The electrons thus produce
a synchrotron 
flux density spectrum 
$S_\nu\propto \nu^{-\alpha}$ with spectral index $\alpha=
s_{\rm p}/2$.  In the original papers \cite{MKB91}
it was assumed that $s_{\rm p}=2$ and correspondingly $\alpha=1$
which is the non-relativistic result from 1st order Fermi acceleration 
at strong shocks in the test-particle approximation.
A flatter value
$s_{\rm p}=3/2$ corresponding to $\alpha=3/4$ may be more
appropriate for strong shocks in the general case \cite{Malkov}.
The cascades are initiated by gamma rays from the decay of the
neutral pions at ultrahigh energies.
The slope
of the differential injection spectrum of the gamma rays 
is given by $\alpha$, but steepens due to
pair creation on the synchrotron target characterized by
the optical depth $\tau(E)\propto E^\alpha$.
The steepening can be described by the energy-dependent
escape probability
$P_{\rm esc}={1-\exp[-\tau(E)]/ \tau(E)}
\rightarrow \tau(E)^{-1}
\propto E^{-\alpha}$ for $\tau\gg 1$, i.e. a 
steepening of the injection spectrum by $\alpha$.  Hence,
the stationary injection spectrum has the slope $s_{\gamma,1}=2\alpha$.
The next step in the cascade development involves the creation
of pairs which have a stationary distribution with the same slope
$2\alpha$ producing the second generation of gamma rays.
Since generally these gamma rays still lie in the optically thick
energy range, the stationary gamma ray distribution has the
slope $s_{\gamma,2}=2\alpha+0.5$.  The same is true for the new generation of
pairs produced by these gamma rays, and their synchrotron gamma rays
are mostly emitted at optically thin energies where their spectrum
has the slope $s_{\gamma,3}=\alpha+0.75$ (in the optically thick range the
slope is $2\alpha+0.75$).  Thus the predicted multi-TeV slope
is bracketed by $2\alpha+0.5$ and $2\alpha+0.75$.  
For $\alpha=1$, the corresponding range is $2.5-2.75$ and for
$\alpha=0.75$ it is $2.0-2.25$ which is in reasonable agreement
with the observations of Mrk~421 and Mrk~501
\cite{Konol99}
corrected for the expected intergalactic gamma ray attenuation
\cite{DJD99}.
The model spectrum fitted to lower frequency data
as published prior to the 1997 HEGRA multi-TeV
observations agrees remarkably well with the measurements
\cite{M98}.
The shape of the multi-TeV spectrum is not sensitive to changes
in the maximum energy and can remain constant under large-amplitude
changes of the flux associated with changes in the maximum energy.

\section{Neutrino and cosmic ray predictions}

\begin{figure*}
\centerline{\psfig{figure=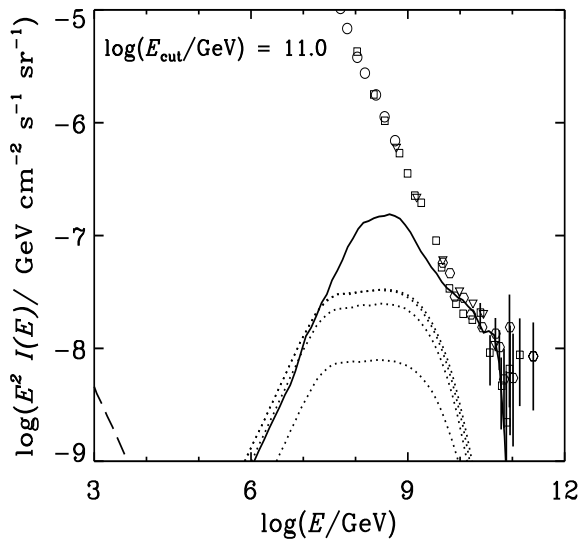,width=10cm,height=8cm}}
\caption{Comparison of proton (solid line) and neutrino fluxes (dotted lines,
from top to bottom $\nu_\mu,~\bar\nu_\mu$, and $\bar\nu_e$) 
from the proton blazar
model 
(Monte-Carlo computations and figure kindly provided by R.J. Protheroe).}
\label{fig3}
\end{figure*}

The photo-production of charged pions leads to the emission of neutrons 
and neutrinos.  
Neutrons associated with the production of $\pi^+$ have no efficient coupling
with the magnetized plasma in the jet and therefore escape ballistically.
The neutrons decay to protons after a propagation length
$l_n=(\gamma_n/10^{11})$~Mpc, and such
extragalactic cosmic rays suffer energy losses traversing the
microwave background \cite{RB93}.  At an observed energy of $10^{19}$~eV,
the energy-loss distance is 
$
\lambda_{\rm p}\sim 1~{\rm Gpc}
$ owing to 
pair production.  This distance corresponds to a redshift $z_{\rm p}$
determined by 
$
\lambda_{\rm p}=(c/ H_\circ)
\int_0^{z_{\rm p}} {dz/[(1+z) E_{\rm JP}(z)]}
$
where 
$
E_{\rm JP}(z)=\left[\Omega(1+z)^3+\Omega_R(1+z)^2+\Omega_\Lambda\right]^{1\over 
2}
$
with $\Omega+\Omega_R+\Omega_\Lambda=1$. 
Almost independent on cosmology, the resulting value for $z_{\rm p}$
is given by 
$
z_{\rm p}= h_{50}/(6-h_{50})\simeq 0.2 h_{50}
$
where $h_{50}=H_\circ/50~$km~s$^{-1}$~Mpc$^{-1}$.
Therefore, when computing the contribution of extragalactic sources
to the observed cosmic ray flux above $10^{19}$~eV, only sources
with $z\le z_{\rm p}$ must be considered. 
Assuming further that extragalactic
sources of cosmic rays and neutrinos
are homogeneously distributed  with a monochromatic
luminosity density
$\Psi(z)\propto (1+z)^{3+k}$
where $k\sim 3$ for AGN \cite{BT98},
their contribution to the energy density of a
present-day diffuse isotropic background is given by
\begin{equation}
u(0)=
\int_0^{z_{\rm m}} \Psi(z) (1+z)^{-4} {dl\over cdz} dz=
{\Psi(0)\over H_\circ} 
\int_0^{z_{\rm m}} {(1+z)^{\rm k}dz\over {(1+z)^2 E_{\rm JP}(z)}}
\end{equation}
where $z_{\rm m}=2$ denotes the redshift of maximum luminosity
density.
The factor
$(1+z)^{-4}$ accounts for the expansion of space and the redshift
of energy. For a simple analytical estimate of the effect
of energy losses on the proton energy density at $10^{19}$~eV,
we collect only protons from sources out to the horizon redshift 
$z_{\rm p}\sim 0.2$
for $10^{19}$~eV protons, whereas neutrinos are collected from sources
out to the redshift of their maximum luminosity density $z_{\rm m}$.
This yields the energy density ratio for neutrinos at an observed
energy of $\sim 5\,10^{17}$~eV and protons at $10^{19}$~eV
\begin{equation}
{u_\nu(0)\over u_{\rm p}(0)}=
{\xi\int_0^{z_{\rm m}} (1+z)^{k-2}/E_{\rm JP}(z) dz\over
\int_0^{z_{\rm p}} (1+z)^{k-2}/E_{\rm JP}(z)dz}\sim 2-3
\end{equation}
using $\xi\sim 0.3$ from decay and interaction kinematics,
and considering 
an open Universe with $E_{\rm JP}(z)=(1+z)$ and a closed
one with $E_{\rm JP}(z)=(1+z)^{3/2}$.    Fig.~\ref{fig3}
shows the exact propagated
proton and neutrino spectra for $\Omega=1$ 
from a full Monte-Carlo
simulation employing the matrix doubling method of 
Protheroe \& Johnson \cite{PJ96}.
The assumed neutron spectrum 
was $dN_n/dE_n\propto E_n^{-1}$ (corresponding to $\alpha=1$
in the previous section) up to $10^{18}$~eV and
$dN_n/dE_n\propto E_n^{-2}\exp(-E_n/E_{\rm cut})$ above.  The
steepening reflects the fact that the maximum energy may
vary from source to source.  
The muon neutrino spectra follow the same shape, but they
are shifted according
to a simplified treatment of
pion decay and production kinematics. 
A more accurate treatment 
yields small corrections 
\cite{MPR99}.
The neutron spectrum was normalized to the cosmic ray data
yielding a neutrino spectrum consistent with 
model A from the original work \cite{M95}.
The associated gamma ray flux $F_\gamma\approx 2F_\nu\ln[100]
\approx 2\times 10^{-6}$~GeV~cm$^{-2}$~s$^{-1}$~sr$^{-1}$
corresponds to a sizable
fraction of the observed gamma ray background flux\footnote{
A recent paper by Waxman and Bahcall \cite{WB99}
refers to the neutrino flux
from model B in the original work which was given only to demonstrate
that hadronic jets can not produce a diffuse gamma ray background
with an MeV bump (as measured by Apollo and which is
now known to be absent from a COMPTEL analysis) without over-producing
cosmic rays at highest energies.}.  The neutrino flux is consistent with
the bound given in ref.~\cite{WB99}, although it is possible to
have extragalactic neutrino sources of higher neutrino fluxes
without violating the observed cosmic ray data as a bound \cite{MPR99}.
One could easily construct such models which produce the entire 
gamma ray background on the same rationale \cite{MPR99}.

\section{Neutrino oscillations and event rates}

The neutrino flux shown in Fig.~\ref{fig3} corresponds to
a very low muon event rate even in a detector with an effective
area of 1~km$^2$
($\sim 1$~event per year and per steradian above 100~TeV).  This event rate
is a very conservative estimate, since there must be
additional neutrino production
due to pp-interactions of escaping nucleons diffusing through
the host galaxies and galaxy clusters.  
The neutrino flux could also be increased
by increasing the number of extragalactic jet 
sources with proton maximum energies
well below $10^{19}$~eV \cite{MPR99}.  As a matter of fact,
such a model ramification is required if one wants to explain
the entire diffuse gamma ray background by hadronic photo-production
sources.

At this point the discovery of neutrino mass announced by the
Super-Kamiokande collaboration 
\cite{SuperK} changes the situation
in a major way.  A deficit of atmospheric muon neutrinos was observed
with Super-Kamiokande at large zenith angles with the most likely
explanation being a full-amplitude oscillation of muon flavor eigenstates to
tauon flavor eigenstates across the Earth at GeV energies.
The transition probability 
$P(\nu_\mu\rightarrow \nu_\tau)$ is a function of distance
and energy $L/E$, i.e.
\begin{equation}
P(\nu_\mu\rightarrow \nu_\tau)\approx \sin^2(2\theta)\sin^2\left(
{\Delta m^2\over 10^{-3}~{\rm eV}^2}{L/10^4~{\rm km}\over E/{\rm GeV}}
\right)
\end{equation}
where the mixing angle $\theta$ specifies the mixing amplitude
$\sin^2(2\theta)>0.82$.
The missing piece of information about the neutrino oscillations is
the appearance of tau leptons to which
Super-Kamiokande is not sensitive. 
A long-baseline experiment using muon neutrinos from a laboratory
beam is extremely difficult if not impossible with existing
laboratories, since one must establish
a very large distance and a high energy to reach the tau mass shell
at $m_\tau=1.784$~GeV.  The beam luminosity decreases rapidly
with energy and distance which poses an irreducible problem.
However, an astrophysical beam of muon neutrinos, such as the one
proposed in this paper and for which the multi-TeV observations
give us somewhat more confidence that they really exist,
is ideally suited for this type of experiment, since both L and E
obtain ``astronomically large'' values.
Hence it follows that very likely astronomical high-energy neutrino sources
such as extragalactic radio jets will help to solve a major puzzle
in elementary particle physics.

There is another effect associated with neutrino oscillations
which alleviates the problem with the low muon event rate
of the predicted flux.  To see this, one must realize that
the solid angle for the detection of high-energy muon neutrinos
becomes very narrow at energies in excess of $\sim 100$~TeV,
since the Earth becomes optically thick to muon neutrinos
above this energy (the weak interaction cross section
depends on energy).  This is called the Earth shadowing effect
\cite{G96}.  Below 100~TeV the atmospheric background of neutrinos
is too strong for the discovery of the rare
events due to extragalactic neutrinos,
unless the angular resolution of the neutrino telescopes is
very good.  Actually, the atmospheric background is comparable
to the predicted astronomical background at around 100~TeV.
Therefore one is confronted with the problem of 
a too small solid angle for events above 100~TeV and a too low
rate below 100~TeV.  If the neutrino oscillation hypothesis is correct,
the extragalactic muon neutrino beam is fully mixed with tau neutrinos
at Earth.  Tau leptons produced by charged-current interactions inside
the Earth 
decay before further interacting.
Since the Earth is also opaque to tau neutrinos above
$\sim$~100~TeV, all the tau neutrinos entering on the one side
with energies above 100~TeV emerge on the other side with energies
of around 100~TeV \cite{HS98}
obliterating the Earth-shadowing effect.
In the proton blazar model with $\alpha=1$, the number density
of neutrinos above 100~TeV remains constant up to roughly $10^5$~TeV
implying an increase in the contained event rate by a factor
\begin{equation}
\xi_\tau={1\over 2}\int_{10^2}^{10^5}{dx\over x}=\ln\left[10^3\right]\simeq 3.5
\end{equation}
where the factor ${1\over 2}$ is for the mixing between muon and tauon
neutrinos (${1\over 3}$ would be appropriate for further mixing with
electron neutrinos).  
The neutrinos from extragalactic radio sources
can therefore be expected to produce  more events than the 
atmosphere at around 100~TeV.  Some of these events are not
muon events, but direct tauon events.  Owing to tauon decay,
the tauon tracks are very short $l_\tau=50(E_\tau/100~\rm TeV)$~cm
and the main signature is the electromagnetic cascade from tau 
decay.  Horizontal events can be of higher energy, producing the
famous double-bang events with the first bang indicating the
charged-current tauon production event and the second its decay.
Thus, the glass is half-full.

\section{Discussion and summary}

The multi-TeV spectra from nearby blazars prediced on the
basis of the proton blazar model
are in accord with the observations if the effect of 
pair attenuation
due to the extragalactic infrared background is taken into account.
Variability patterns are similar to those in synchrotron-self-Compton
models, but more complex,
since variations of the target photon flux fold into the cascade
development in a non-linear manner. Short time scale variability is
likely to reflect the passage of shocks through inhomogeneities
and correspond to cooling time scale variations.
If the
cosmic ray flux emitted by hadronic accelerators is enough
to explain the observed
cosmic rays above $10^{19}$~eV, the associated gamma ray power
from these sources is enough to produce 
at least a sizable fraction of the observed extragalactic
gamma ray background. 
The gamma ray power is larger than that in cosmic rays, since
the cosmic rays lose energy traversing the low-energy background
radiation fields and most sources have high redshifts.
Strong evolution of their luminosity density would rule out
GRBs as possible sources of the highest energy cosmic rays,
since their cumulative gamma ray flux is far below the extragalactic
gamma ray flux.
Although the expected muon event rate in neutrino telescopes is low, 
neutrino oscillations lead to tau cascades canceling the Earth shadowing
effect thereby increasing the detection probability.

\end{document}